\documentclass[preprint,prl,nofootinbib]{revtex4}
\usepackage{graphicx}
\usepackage{dcolumn}
\usepackage{amsmath,amsthm,amssymb}
\usepackage[active]{srcltx} 
\usepackage{color}
\usepackage{fancybox}

\def\d{\partial}

\def\Re{\text{Re\,}}
\def\Im{\text{Im\,}}
\def\la{\lambda}
\def\Res{\mathop{\text{Res}}}

\def\ra{\rightarrow}
\providecommand{\norm}[1]{\lVert#1\rVert}

\begin{document}

\title{Dynamics near the threshold for blowup in the one-dimensional focusing nonlinear Klein-Gordon equation}

\author{Piotr Bizo\'n}
\affiliation{M. Smoluchowski Institute of Physics, Jagiellonian University, Krak\'ow, Poland}
\author{Tadeusz Chmaj}
\affiliation{H. Niewodniczanski Institute of Nuclear
   Physics, Polish Academy of Sciences,  Krak\'ow, Poland}
   \affiliation{Cracow University of Technology, Krak\'ow,
    Poland}
\author{Nikodem Szpak}
\affiliation{Faculty of Physics, University Duisburg-Essen, Germany}

\date{\today}
\begin{abstract}
We study dynamics near the threshold for blowup in the focusing nonlinear Klein-Gordon equation $u_{tt}-u_{xx} +  u - |u|^{2\alpha} u =0$ on the line. Using mixed numerical and analytical methods we find that solutions starting from even initial data, fine-tuned to the threshold, are trapped by the static solution $S$ for intermediate times. The details of trapping are shown to depend on the power $\alpha$, namely, we observe fast convergence to $S$ for $\alpha>1$, slow convergence for $\alpha=1$, and very slow (if any) convergence for $0<\alpha<1$.
 Our findings are complementary with respect to the recent rigorous analysis of the same problem (for $\alpha>2$) by Krieger, Nakanishi, and Schlag \cite{kns}.
\end{abstract}

\maketitle

\section{Introduction}

In this paper we consider the nonlinear Klein-Gordon equation (NLKG) with a focusing power nonlinearity in one spatial dimension
\begin{equation}\label{main}
u_{tt}-u_{xx} +  u - |u|^{2\alpha} u =0\,,\qquad \alpha>0\,,
\end{equation}
for real-valued $u(t,x)$ and smooth compactly supported (or exponentially localized) even initial data. This equation arises in a variety of physical applications, for instance as a model of self-focusing waves in nonlinear optics.
It
is well known that for some initial data solutions of Eq.\eqref{main} are globally regular in time and for some initial data they  blow up in  finite time. In particular, the global regularity
for small data follows from standard energy estimates \cite{c}, while blowup for negative energy data follows from a concavity argument \cite{l}. In addition, not all globally regular solutions decay to zero, as is evident from the fact that there exists the (unstable) static solution
\begin{equation}\label{static}
u(t,x)=S(x)=\frac{(\alpha+1)^{\frac{1}{2\alpha}}}{(\cosh{\alpha x})^{1/\alpha}}\,.
\end{equation}

The dichotomy of global regularity vs blowup raises two obvious questions: (i) what determines
a borderline between these two behaviours and (ii) what is the evolution of critical
initial data which lie on the borderline.
These questions have been recently studied by Krieger, Nakanishi, and Schlag \cite{kns}. They proved that there exist a  codimension-one center-stable manifold associated with $S$ which locally separates the regions of blowup and dispersion, and solutions lying on this manifold are trapped by $S$ (i.e., remain in its small neighborhood) for all future times. Moreover, they proved that for $\alpha>2$ trapping actually implies  asymptotic convergence to $S$, however the rate of convergence was not determined.

 The aim of this paper is to obtain a precise quantitative description of the asymptotic dynamics on the center-stable manifold (decay rates and spatial profiles). To this end, we proceed in two steps using a combination of  analytical and numerical methods. In step one, we derive analytically the asymptotic behaviour of linearized perturbations about $S$.
 In step two,
  we employ the numerical bisection method to fine-tune
 initial data to
the borderline between blowup and global regularity. Such specially prepared solutions  approach $S$, hang around it for some time (whose span depends on the accuracy of fine-tuning), and eventually are ejected out along the one-dimensional unstable manifold of $S$. We find that this behaviour is very well approximated by the linearized dynamics obtained in step one. The main outcome of our study is
that the dynamics on the center-stable manifold of $S$ exhibits an interesting dependence on the power $\alpha$ which is a consequence of the change of spectral properties of the linearized operator around $S$ as $\alpha$ varies. More precisely, we distinguish three different scenarios of convergence to $S$ as $t\ra\infty$: (a) fast dispersive decay $\norm{u(t,x)-S(x)}_{\infty}\sim  t^{-3/2} \sin(t)$ for $\alpha>1$, (b)  slow dispersive decay $\norm{u(t,x)-S(x)}_{\infty}\sim  t^{-1/2}\sin(t)$  for $\alpha=1$ due to the presence of the zero energy resonance, and (c) very slow (if any) decay
for $0<\alpha<1$ due to the presence of oscillatory modes.

 The paper is organized as follows. In section~2 we recall some well-known facts about the spectrum of the linearized operator around $S$. These facts are used in section~3 to describe the linearized dynamics near $S$. In section~4 we present  numerical simulations of the critical dynamics for Eq.\eqref{main} for three representative powers, $\alpha=\frac{1}{2},1$, and $\frac{3}{2}$, and confront them with the results of section~3. Finally, in section~5 we mention  several open problems for future work.

 \section{Spectrum of the linearized operator}
In order to make the paper self-contained and to fix notation, in this section we recall the well-known facts about the spectrum of the linearized operator around the static solution~$S$.

Substituting $u(t,x)=S(x)+f(t,x)$ into Eq.\eqref{main} and linearizing, we obtain the linear evolution equation for small perturbations around $S$
\begin{equation}\label{weq}
f_{tt}-f_{xx} +  f + V(x) f  =0\,,\qquad V(x)=-\frac{(2\alpha+1)(\alpha+1)}{\cosh^2{\!\alpha x}}\,.
\end{equation}
After separation of variables, $f(t,x)=e^{st} v(x)$,  we get the eigenvalue
problem
\begin{equation}\label{pt}
L v = -\lambda^2 v\,,\qquad L=-\frac{d^2}{dx^2} + V(x)\,,
\end{equation}
where $\lambda^2=s^2+1$.
The potential $V(x)$ is the exactly solvable P\"oschl-Teller potential. Since we are interested only in even eigenfunctions, we restrict the domain of the operator $L$ to square-integrable functions on the positive half-line with the Neumann boundary condition $v'(0)=0$.
Then $L$ is self-adjoint with the continuous spectrum $[0,\infty)$ and a discrete spectrum depending on $\alpha$. The generalized eigenfunctions  have the form
\begin{equation}\label{geneig}
    v^{(\alpha)}(\la,x)=\left(\cosh{\alpha x}\right)^{(2+1/\alpha)}
{}_2F_1\left(1+\frac{1-\lambda}{2\alpha},1+\frac{1+\lambda}{2\alpha},\frac{1}{2};-\sinh^2{\!\alpha x}\right)\,.
\end{equation}
Using the connection formula for hypergeometric functions (for a non-integer $a-b$)
\begin{eqnarray}\label{Fas}
    {}_2F_1(a,b,c;z)&=& \frac{\Gamma(c)\Gamma(b-a)}{\Gamma(b)\Gamma(c-a)}\, (1-z)^{-a}
     {}_2F_1\left(a,c-b,1+a-b;\frac{1}{1-z}\right)\nonumber\\
    & +& \frac{\Gamma(c)\Gamma(a-b)}{\Gamma(a)\Gamma(c-b)}\, (1-z)^{-b}
     {}_2F_1\left(b,c-a,1+b-a;\frac{1}{1-z}\right)\,,
\end{eqnarray}
we rewrite \eqref{geneig} in the form (for $\la\neq 0$)
\begin{equation}\label{Fasym}
    v^{(\alpha)}(x,\la) = A_{\alpha}(\lambda) v_{-}^{(\alpha)}(\lambda,x) + A_{\alpha}(-\lambda) v_{+}^{(\alpha)}(\lambda,x)\,,
\end{equation}
where
\begin{eqnarray}\label{A}
    A_{\alpha}(\lambda)&=&\frac{\Gamma\left(\frac{1}{2}\right)\Gamma\left(\frac{\lambda}{\alpha}\right)}
    {\Gamma\left(1+\frac{1+\lambda}{2\alpha}\right)
    \Gamma\left(-\frac{1}{2}-\frac{1-\lambda}{2\alpha}\right)}\,,\\
    v_{\pm}^{(\alpha)}(\lambda,x)&=&(\cosh{\alpha x})^{\mp\lambda/\alpha} {}_2F_1\left(1+\frac{1\pm\lambda}{2\alpha},-\frac{1}{2}-
    \frac{1\mp\lambda}{2\alpha},1\pm\frac{\lambda}{\alpha};\cosh^{-2}{\!\alpha x}\right)\,.
\end{eqnarray}
 For $x\rightarrow\infty$ we have
  \begin{equation}\label{in/out}
  v_{\pm}^{(\alpha)}(\lambda,x)\sim 2^{\pm \la/\alpha} e^{\mp \lambda x}\,,
   \end{equation}
   so $v_{-}^{(\alpha)}(\la,x)$ and $v_{+}^{(\alpha)}(\la,x)$ represent ingoing and outgoing waves, respectively.
Eigenvalues are given by the positive roots of the equation $A_{\alpha}(\lambda)=0$:
  \begin{equation}\label{roots}
  \lambda_n(\alpha)=\alpha+1-2n\alpha\,,\qquad n=0,1,\dots N\,,
    \end{equation}
    where $N$ is the largest integer less than $(\alpha+1)/2\alpha$.
Note that due to the dispersion relation $\lambda^2=s^2+1$,  eigenvalues with $\lambda>1$ correspond to unstable modes which grow exponentially in time as $e^{\sqrt{\la^2-1}\,t}$, while eigenvalues lying in the interval $0<\la<1$ correspond to oscillatory (neutral) modes behaving as $\sin(\sqrt{1-\la^2}\,t)$. Thus, for each $\alpha$ there is exactly one unstable eigenmode (with the eigenvalue $\lambda_0(\alpha)=\alpha+1$), while for $\alpha<1$ there are, in addition, $N$ oscillatory eigenmodes.
Denoting by $v_n^{(\alpha)}(x)$ an eigenfunction associated with the eigenvalue $\lambda_n(\alpha)$ we get from \eqref{Fasym}, (10), and \eqref{roots}
\begin{eqnarray}\label{eigfun}
    v^{(\alpha)}_n(x)&:=&v^{(\alpha)}(\lambda_n,x)=A_{\alpha}(-\lambda_n) v_{+}^{(\alpha)}(\la_n,x)\\ \nonumber
   &=& \left(\cosh{\alpha x}\right)^{-(1+1/\alpha)+2n}\,
{}_2F_1\left(\frac{3}{2}-n+\frac{1}{\alpha},-n,2-2n+\frac{1}{\alpha};\cosh^{-2}{\!\alpha x}\right)\,.
\end{eqnarray}
The hypergeometric function above is a polynomial in $\cosh^{-2}\!{\alpha x}$ of order $n$. Factorizing this polynomial we get
\begin{equation}
v_n^{(\alpha)}(x)=\left(\cosh{\alpha x}\right)^{-(1+1/\alpha)}\prod_{k=1}^{n}\left(1-c_{nk}\sinh^2\!{\alpha x}\right)\,,
\end{equation}
with certain positive coefficients $c_{nk}$.
For example, the first two eigenfunctions are
 \begin{eqnarray}\label{ground}
    v_0^{(\alpha)}(x)&=&\left(\cosh{\alpha x}\right)^{-(1+1/\alpha)}\,,\hspace{4cm} \lambda_0^{(\alpha)}=\alpha+1\,,\\
    v_1^{(\alpha)}(x)&=&\left(\cosh{\alpha x}\right)^{-(1+1/\alpha)}\left(1-\frac{2}{\alpha}\sinh^2\!{\alpha x}\right)\,,\qquad  \lambda_1^{(\alpha)}=1-\alpha\,.
\end{eqnarray}

An important role in our analysis will be played by the generalized eigenfunction at the endpoint of the continuous spectrum $\lambda=0$
\begin{equation}\label{bottom}
    v^{(\alpha)}(0,x)=\left(\cosh{\alpha x}\right)^{(2+1/\alpha)} {}_2F_1\left(1+\frac{1}{2\alpha},1+\frac{1}{2\alpha},\frac{1}{2};-\sinh^2{\!\alpha x}\right)\,.
\end{equation}
Using the formula (see 15.8.9 in \cite{nist})
\begin{eqnarray}\label{Faa}
    {}_2F_1(a,a,c;z)&=& \frac{\Gamma(c)(1-z)^{-a}}{\Gamma(a)\Gamma(c-a)}\, \sum_{k=0}^{\infty}\frac{(a)_k(c-a)_k}{(k!)^2}\,(1-z)^{-k} \\ \nonumber
    && \times \left(\ln(1-z)+2\psi(k+1)-\psi(a+k)-\psi(c-a+k)\right)\,,
\end{eqnarray}
we obtain the asymptotic behaviour for $x\rightarrow\infty$
\begin{equation}\label{zeroasym}
v^{(\alpha)}(0,x)=\frac{2\alpha\sqrt{\pi}}{\Gamma(1+\frac{1}{2\alpha})
\Gamma(-\frac{1}{2}-\frac{1}{2\alpha})}\, x +\mathcal{O}(1)\,.
\end{equation}
 If $(\alpha+1)/2\alpha$ is a positive integer (i.e., $\alpha=1,1/3,1/5,...$), then the coefficient of the leading order term in \eqref{zeroasym} vanishes and therefore $v^{(\alpha)}(0,x)$ is everywhere bounded,
 which means that there is a resonance at the endpoint of the continuous spectrum (this is also seen as the zero "eigenvalue" in \eqref{roots}).
 \vskip 0.2cm
 \noindent \emph{Remark 1.}   Note that negative zeros of the coefficient $A_{\alpha}(\la)$ correspond to antibound states. For $\alpha\neq 1/k$ ($k\in\mathbb{N}$) they are given by two infinite series
    \begin{equation}\label{reson}
    \lambda_n^{-}(\alpha)=\alpha+1-2n\alpha\,\quad\text{and}\quad     \lambda_m^{+}(\alpha)=-(2\alpha+1+2m\alpha)\,,
    \end{equation}
    where $n$ runs over all integers greater than  $(\alpha+1)/2\alpha$ and $m$ runs over all non-negative integers. For $\alpha=1/k$ there are no antibound states because  of the cancelation of infinities in the gamma functions in the numerator and the denominator of $A_{\alpha}(\lambda)$.
  \vskip 0.2cm
 \noindent \emph{Remark 2.} In two important cases of quadratic and cubic nonlinearities the generalized eigenfunctions $v^{(\alpha)}(\la,x)$  can be expressed by elementary functions. For $\alpha=1/2$ we have
\begin{eqnarray}\label{alpha1/2}
(\la^2-1)  v^{(\frac{1}{2})}(\la,x)&=&\cosh(\la x) \left(\la^2+\frac{11}{4}-\frac{15} {4 \cosh^2{\!\frac{x}{2}}}\right)\\
&-& \frac{\sinh(\la x)}{\la} \tanh\left(\frac{x}{2}\right) \left(3\la^2 +\frac{3}{4} -\frac{15}{8 \cosh^2{\!\frac{x}{2}}}\right)\,,
\end{eqnarray}
and for $\alpha=1$ we have
\begin{equation}\label{al1}
  (\la^2-1)  v^{(1)}(\la,x)=\cosh(\la x) \left(\la^2+2-\frac{3}{\cosh^2{\!x}}\right)-3\la\tanh(x) \sinh(\la x)\,.
\end{equation}

\section{Linearized dynamics}

We will show in the next section that one can prepare special initial data for which the solution approaches the static solution $S$ and remains close to it for some time. It is natural to expect that during this transient phase of evolution the dynamics can be approximated by the linearization around $S$ which leads us to the study of Eq.\eqref{weq}.
For $t\geq 0$  the solutions of this linear equation
  are given by
\begin{equation}
  f(t,x) = \int \partial_t G(t,x,y) f(0,y) dy + \int G(t,x,y) \partial_t f(0,y) dy\,,
\end{equation}
where $G(t,x,y)$ is the retarded Green function.
In order to determine the asymptotic behaviour of Green's function for late times we shall use the Laplace transform in time
\begin{equation}
  g(s,x,y) := \int_0^\infty e^{-st} G(t,x,y) dt\,.
\end{equation}
which  is well-defined and
 analytic in $s$ for $\Re{s}>\sqrt{\alpha(2+\alpha)}$.
Following the standard procedure, we construct $g(s,x,y)$ from two linearly independent solutions\footnote{Hereafter, to avoid notational clutter we drop the superscript $(\alpha)$ on the eigenfunctions.} of Eq.\eqref{pt}, $v(\la,x)$ and $v_{+}(\la,x)$, satisfying the appropriate boundary conditions: $v'(\la,0)=0$ and $v_{+}(\la,x)\sim e^{-\la x}$ for $x\ra\infty$,
\begin{equation}\label{green2}
  g(s,x,y)=\hat g(\la,x,y) := \frac1{W(\la)}\left\{\begin{array}{ll}
    v(\la,x)\, v_{+}(\la,y) & \text{for }x\leq y \\
    v(\la,y)\, v_{+}(\la,x) & \text{for }y<x
  \end{array} \right.
\end{equation}
where $W(\la)$ is the Wronskian of solutions $v(\la,x)$ and $v_{+}(\la,x)$. Taking advantage of the fact that the Wronskian does not depend on $x$ and computing it at infinity using \eqref{Fasym} and \eqref{in/out} we get
$W(\la)=-2\la A(\la)$.

The function $\hat g(\la,x,y)$  is analytic in $\la$ for
$\Re(\la)>\la_0$ and  has a meromorphic continuation
to the whole complex plane of $\la$ with a countable number of poles located at
 the zeros of the Wronskian.
The function $g(s,x,y)$ inherits all these poles through the dispersion relation $\la=\sqrt{s^2+1}$ and, in addition,  has two branch points at $s=\pm i$.
We take the branch cuts emanating from the points $s=\pm i$ and going horizontally to the left  ($\Im(s)=\pm 1$ and $\Re(s)\ra -\infty$). This choice is convenient because
 only the poles $s_n$ corresponding to the eigenvalues $\sqrt{s_n^2+1}=\la_n>0$ given by \eqref{roots}  lie on the first Riemann sheet while the poles corresponding to the antibound states $\la^\pm_n<0$ given by \eqref{reson}  lie on the second sheet\footnote{With this choice of cuts, for a pole $\la>1$ there are two poles $s=\pm\sqrt{\la^2-1}$ on the first sheet. As $\la$ decreases and crosses the value $+1$ these poles meet at $s=0$ and continue as a pair $s=\pm i\sqrt{1-\la^2}$. For $\la=0$ they reach the branch points $\pm i$. Decreasing $\la$ below zero causes the two poles go back to $s=\pm i\sqrt{1-\la^2}$ and further to $s=\pm\sqrt{\la^2-1}$ but now on the second sheet.}.

Next, having $g(s,x,y)$, we compute $G(t,x,y)$ via the inverse Laplace transform
\begin{equation} \label{G-invlap}
  G(t,x,y) = \frac1{2\pi i} \int_{c-i\infty}^{c+i\infty} g(s,x,y) e^{st} ds\,,
\end{equation}
where $c$ is taken to the right of all
the singularities, which in our case means $c>\sqrt{\alpha(2+\alpha)}$.
In order to obtain the late-time asymptotics  $t\gg |x-y|$ of $G(t,x,y)$, we deform the contour of integration  to the left half-plane as shown in Fig.~1
and use the Cauchy residue theorem.
\begin{figure}[ht]
\center
\includegraphics[width=0.4\textwidth]{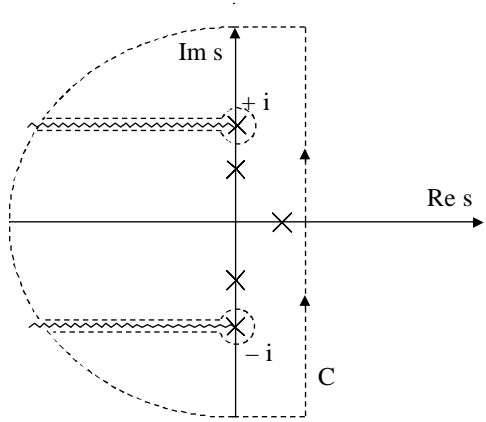}
\caption{The deformation of the contour of integration.
\label{fig:p1}}
\end{figure}
\vskip -0.3cm
\noindent For late times $t\gg |x-y|$ the integral over the semi-circle at infinity vanishes
so $G=G_R+G_C$, where  $G_R$ is the contribution from  the sum over residua  at the poles of $g(s,x,y)$ lying on the first sheet
and $G_C$ is the contribution from the integrals surrounding the two branch cuts.
We have
\begin{equation}
  G_R=\sum_{n=0}^N \Res_{s=s_n}[g(s,x,y) e^{st}] = a_0 e^{s_0 t} v_0(x) v_0(y) + \sum_{n=1}^N a_n  \sin(\omega_n t+\delta_n) v_n(x) v_n(y)\,,
\end{equation}
where $s_0=\sqrt{\alpha(\alpha+2)}$ and $\omega_n:=\sqrt{1-\la_n^2}$ with $\la_n$ given in \eqref{roots}.

The integrals along the branch cuts, because of the exponential damping $e^{st}$ for $\Re(s)<0$, are dominated by contributions from  the endpoints  $s=\pm i$. The behaviour of $g(s,x,y)$ near the endpoints follows from the Laurent expansion of $\hat g(\la,x,y)$ at $\la=0$
\begin{equation}\label{ghat}
    \hat g(\la,x,y)=\frac{c_{-1}(x,y)}{\la}+c_0(x,y)+c_1(x,y) \la + \mathcal{O}(\la^2)\,,
\end{equation}
where $c_{-1}$ is nonzero if and only if  there a resonance at zero energy.
Let $\Delta^{\pm}g(s)$ be the jump of $g(s,x,y)$ across the upper/lower cut. From \eqref{ghat} we get
$\Delta^{\pm} g(s) \sim (s\mp i)^{\beta}$ near $s=\pm i$, where $\beta=-1/2$ if there is a zero energy resonance and otherwise $\beta=1/2$. This yields the late-time asymptotic behaviour
\begin{equation}\label{GC}
    G_C=-\frac1{2\pi i} \int_{cuts}  g(s,x,y) e^{st} ds \sim
    \begin{cases} t^{-1/2} \sin(t)\, c_{-1}(x,y) & \text{resonance at zero,}
\\
 t^{-3/2} \sin(t)\, c_1(x,y) & \text{no resonance at zero.}
\end{cases}
\end{equation}
It is evident from the expression \eqref{green2} that $c_{-1}(x,y)\sim v(0,x) v(0,y)$. Somewhat surprisingly, if zero is not a resonance, the same is true for $c_1(x,y)$. To see this, let us differentiate Eq.\eqref{pt} with respect to $\la$ and take the limit $\la\ra 0$. We get $L (\d_\la \hat g)(0,x,y) = -\lim_{\la\ra 0} 2\la \hat g(\la,x,y) = 0$, hence $\d_\la \hat g(0,x,y)$ is a zero mode of $L$. Since by construction $\d_\la \hat g(0,x,y)$ satisfies the Neumann boundary condition at zero and  is symmetric in $x,y$, it follows that $c_1(x,y)\sim v(0,x) v(0,y)$, as claimed.

\section{Numerical results}

In this section we present numerical simulations of the critical behaviour for Eq.\eqref{main} on the half-line $x\geq 0$ with the Neumann boundary condition $u_t(t,0)=0$ and exponentially localized initial data. The numerical technique is standard. We use the method of lines with a fourth-order Runge-Kutta time integration and  fourth-order spatial finite differences. All simulations were performed in the quadruple (128-bit) arithmetic  precision.

Before presenting the results let us recall that Eq.\eqref{main} has the conserved energy
\begin{equation}\label{E}
E(u,u_t)=\frac{1}{2} \int_0^{\infty} \left(u_t^2 +u_x^2 + u^2-\frac{1}{\alpha+1}\, |u|^{2\alpha+2}\right) dx\,.
\end{equation}
The static solution $S$ is the critical point of the static energy functional $E_0(u):=E(u,0)$. Another useful functional is
\begin{equation}\label{K}
K(u):=\frac{d}{d\beta} E_0(\beta u)|_{\beta=1}=\int_0^{\infty} \left(u_x^2 + u^2-|u|^{2\alpha+2}\right) dx\,.
\end{equation}
Since $\delta E_0(S)=0$, it follows that $K(S)=0$.
The significance of the functional $K$ was first pointed out by Payne and Sattinger \cite{ps}
 who showed that for solutions with $E(u,u_t)<E_0(S)$ one has dichotomy: global regularity
 for $K(u)\geq 0$ and blowup for $K(u)<0$.
 Note that the boundaries of the Payne-Sattinger sets $\mathcal{K}_+=\{(u,u_t)|E(u,u_t)<E_0(S), K(u)\geq 0\}$ and
$\mathcal{K}_-=\{(u,u_t)|E(u,u_t)<E_0(S), K(u)< 0\}$ are separated (except for a single point $(S,0)$),
 hence the borderline between global regularity and blowup necessarily lies above $E_0(S)$.

The results shown below correspond to the following one-parameter family of initial data\footnote{We stress that the near-critical dynamics is universal in the sense that it does not depend on a particular choice of a family of initial data which interpolates between basins of global regularity and blowup.}
\begin{equation}\label{idata}
    u(0,x)=(\alpha+1)^{\frac{1}{2\alpha}} \exp\left(-\frac{x^2}{\sigma^2}\right)\,, \qquad u_t(0,x)=0\,,
\end{equation}
Note that these data lie entirely above the energy $E_0(S)$. We find that solutions with small $\sigma$ are globally regular while
 solutions with large $\sigma$ blow up in finite time. Using bisection we find that there is a single $\sigma^*$ which separates these two regions.

We remark in passing that by repeating the bisection for many different families of initial data (depending on one, two, or even three parameters)
 one can probe the shape of the center-stable manifold in the phase space. Recently,  very interesting studies in this direction were reported in \cite{ds}, however we do not pursue them here since
 our goal is different. We want to get the precise quantitative description of the critical evolution. For this purpose,  as emphasized above, the choice of an interpolating family of initial data is irrelevant.

At $\sigma^*$ the curve of initial data \eqref{idata} intersects the center-stable manifold of $S$. Of course, numerically it is impossible to start the evolution exactly at this intersection point, so hereafter we show only
near-critical solutions (with $\sigma^*$ determined to about 30 decimal places). Such solutions, by continuity, approach $S$ and stay in its neighborhood for some time until they are ejected out along the one-dimensional unstable manifold of $S$.
 On the basis of linearized analysis from section~3 we expect that when the near-critical solution stays close to $S$ it should  be well approximated by the expression
\begin{equation}\label{lin_expansion}
u(t,x)\simeq S(x)+ A_0\, e^{s_0 t} v_0(x) +  \sum_{n=1}^{N} A_n \sin(\omega_n t+\delta_n) \,v_n(x)
+C\, t^{-\gamma}\,\sin(t) \,v(0,x)\,,
\end{equation}
where $s_0=\sqrt{\alpha(\alpha+2)}$, $\omega_n=\sqrt{1-(\la_n)^2}$, and $\gamma=1/2$ (if there is a zero energy resonance) or $\gamma=3/2$ (otherwise). For  critical data the coefficient $A_0(\sigma^*)$ vanishes, however in practice there is always a small admixture of the unstable mode which means that the length of time during which we can observe the near-critical dynamics is limited by a characteristic time of growth of the instability $\sim -\frac{1}{s_0} \ln|\delta\sigma|$, where $\delta\sigma$ is the distance from the threshold achieved by bisection.
\addtolength{\topmargin}{-2pc}
\addtolength{\textheight}{4pc}

The results of our numerical simulations are depicted  in Fig.~2,3,4 for three  powers $\alpha=1/2,1$, and $3/2$ (corresponding to
quadratic, cubic, and quartic nonlinearities, respectively). They illustrate three qualitatively different scenarios of critical dynamics.
 In each case
 we show two plots: (a)  time evolution at a fixed point in space (for convenience we take $x=0$) for a pair of marginally sub- and super-critical initial data \eqref{idata} with $\sigma=\sigma^*\pm 10^{-30}$, and (b) snapshots of spatial profiles at several times before ejection. Below, in order to facilitate the comparison of numerics with the predictions of linear theory, we give the explicit form of the expression \eqref{lin_expansion} specialized to each case.
\begin{itemize}
\item $\alpha=3/2$: The linear approximation \eqref{lin_expansion} takes the form
\begin{equation}\label{alpha32}
u(t,x)\simeq S(x) + A_0\, e^{\frac{\sqrt{15}}{2}\, t} v_0(x) +  C \,t^{-3/2} \sin(t+\delta)  v(0,x)\,,
\end{equation}
where
\begin{eqnarray}\label{v3/2}
S(x)&=& \left(\frac{5}{2}\right)^{\frac{1}{3}}\left(\cosh\frac{3x}{2}\right)^{-\frac{2}{3}}\,,\\
    v_0(x)&=&\left(\cosh\frac{3x}{2}\right)^{-\frac{5}{3}}\,,\\
    v(0,x)&=& \left(\cosh\frac{3x}{2}\right)^{\frac{8}{3}} {}_2F_1\left(\frac{4}{3},\frac{4}{3},\frac{1}{2};-\sinh^2\frac{3x}{2}\right)\,.
\end{eqnarray}
\addtolength{\topmargin}{+2pc}
\addtolength{\textheight}{-4pc}
\begin{figure}[h]
\center
\includegraphics[width=0.49\textwidth]{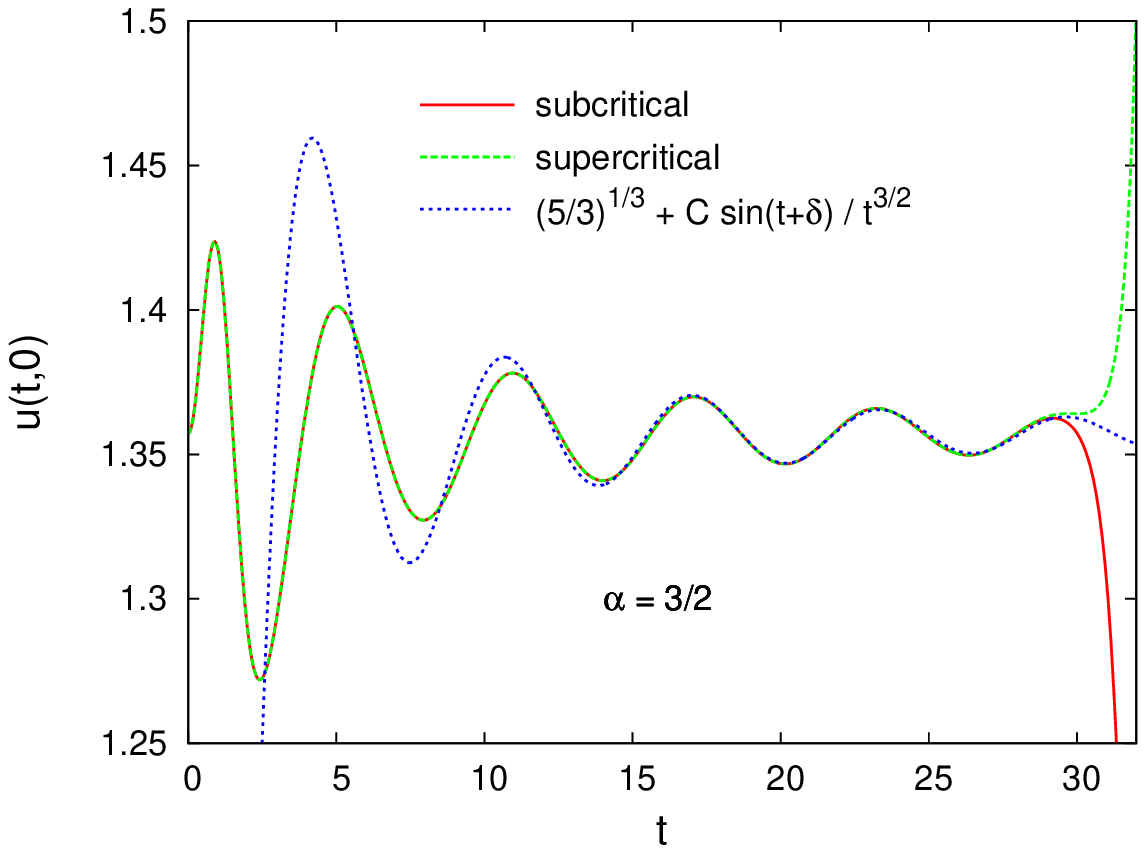}
\includegraphics[width=0.49\textwidth]{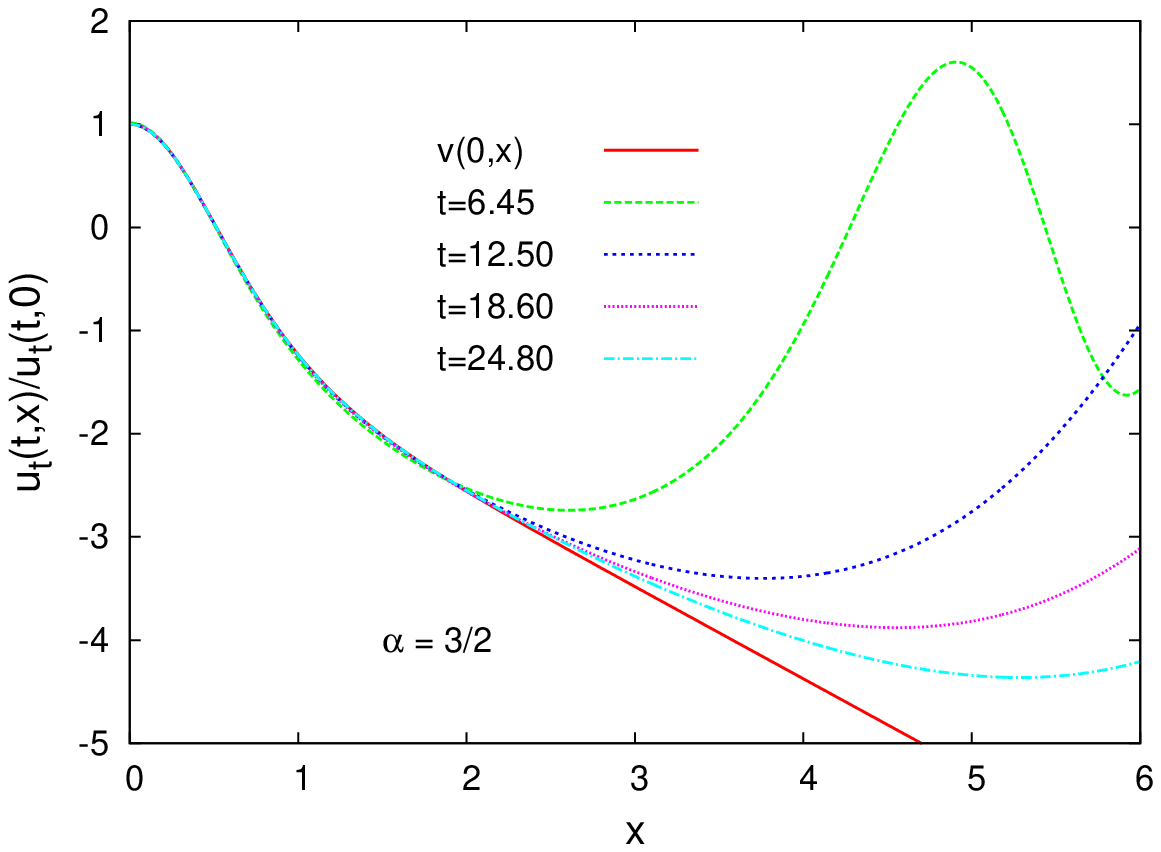}
\caption{$\alpha=3/2$. (a) Fast dispersive decay to $S$, (b) Convergence of the spatial profile to the zero energy eigenfunction (37). The critical behaviour for all $\alpha>1$ is qualitatively the same.
\label{fig:p2}}
\end{figure}
\item $\alpha=1$: The linear approximation \eqref{lin_expansion}  takes the form
\begin{equation}\label{alpha1}
u(t,x)\simeq S(x)+ A_0\, e^{\sqrt{3}\, t} v_0(x) +  C\, t^{-1/2} \sin(t+\delta) \, v(0,x)\,,
\end{equation}
where
 \begin{eqnarray}\label{v1}
 S(x)&=&\sqrt{2} \cosh^{-1}(x)\,,\\
    v_0(x)&=&\cosh^{-2}(x)\,,\\
    v(0,x)&=& 3\cosh^{-2}(x)-2\,.
\end{eqnarray}
\begin{figure}[h]
\center
\includegraphics[width=0.49\textwidth]{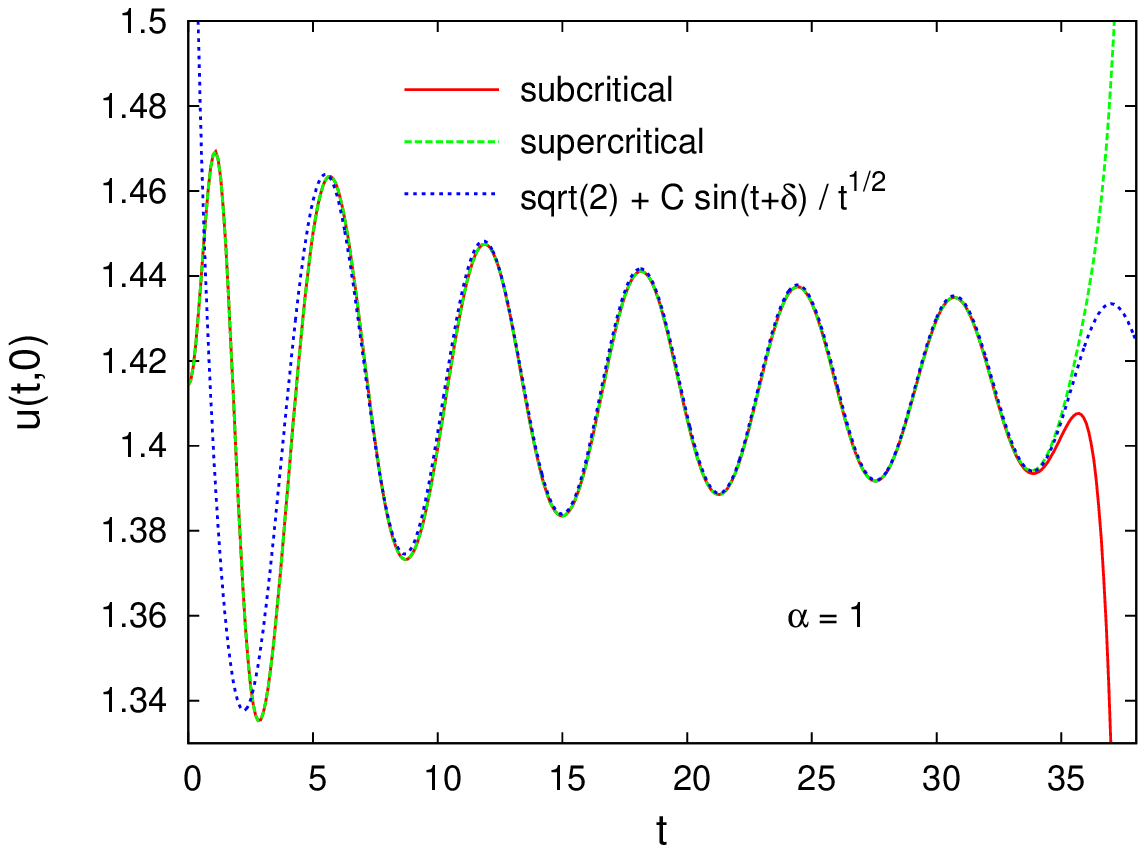}
\includegraphics[width=0.49\textwidth]{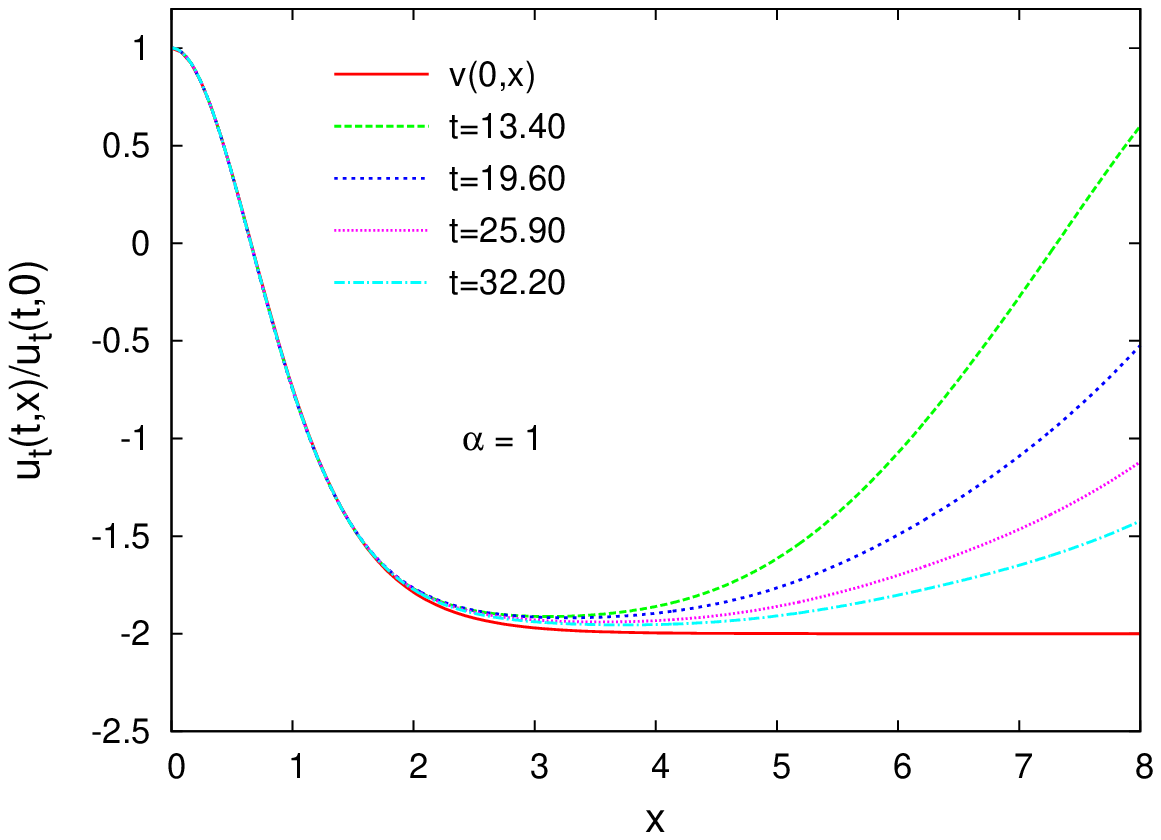}
\caption{$\alpha=1$. (a) Slow dispersive decay to $S$, (b) Convergence of the spatial profile to the zero energy (resonance) eigenfunction (41).
\label{fig:p3}}
\end{figure}
\item $\alpha=1/2$: The linear approximation \eqref{lin_expansion}  takes the form
\begin{equation}\label{alpha05}
u(t,x)\simeq S(x)+ A_0\, e^{\frac{\sqrt{5}}{2}\, t} v_0(x) +  A_1 \sin(\tfrac{\sqrt{3}}{2} t+\delta_1)\, v_1(x)+ C \,t^{-3/2} \sin(t+\delta) \, v(0,x)\,,
\end{equation}
where
 \begin{eqnarray}\label{v05}
 S(x)&=&\frac{3}{2} \left(\cosh\frac{x}{2}\right)^{-2}\,,\\
    v_0(x)&=&\left(\cosh\frac{x}{2}\right)^{-3}\,,\\
    v_1(x)&=&\left(\cosh\frac{x}{2}\right)^{-3}
    \left(1-4\sinh^2\frac{x}{2}\right)\,,\\
    v(0,x)&=& \frac{15}{4} \left(\cosh\frac{x}{2}\right)^{-2} -\frac{11}{4}+
 x \tanh\left(\frac{x}{2}\right) \left(\frac{3}{4} -\frac{15}{8} \left(\cosh\frac{x}{2}\right)^{-2}\right)\,.
\end{eqnarray}
\addtolength{\topmargin}{-2pc}
\addtolength{\textheight}{4pc}
\begin{figure}[h]
\center
\includegraphics[width=0.49\textwidth]{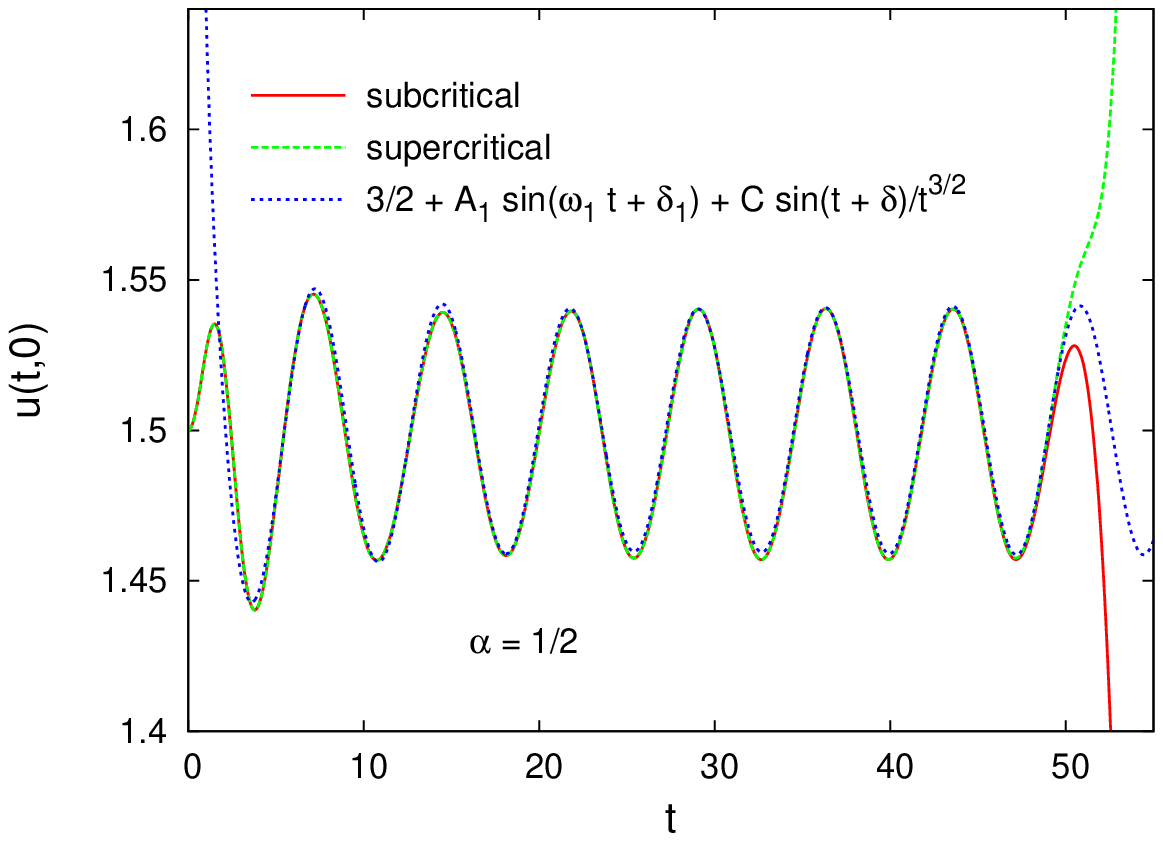}
\includegraphics[width=0.49\textwidth]{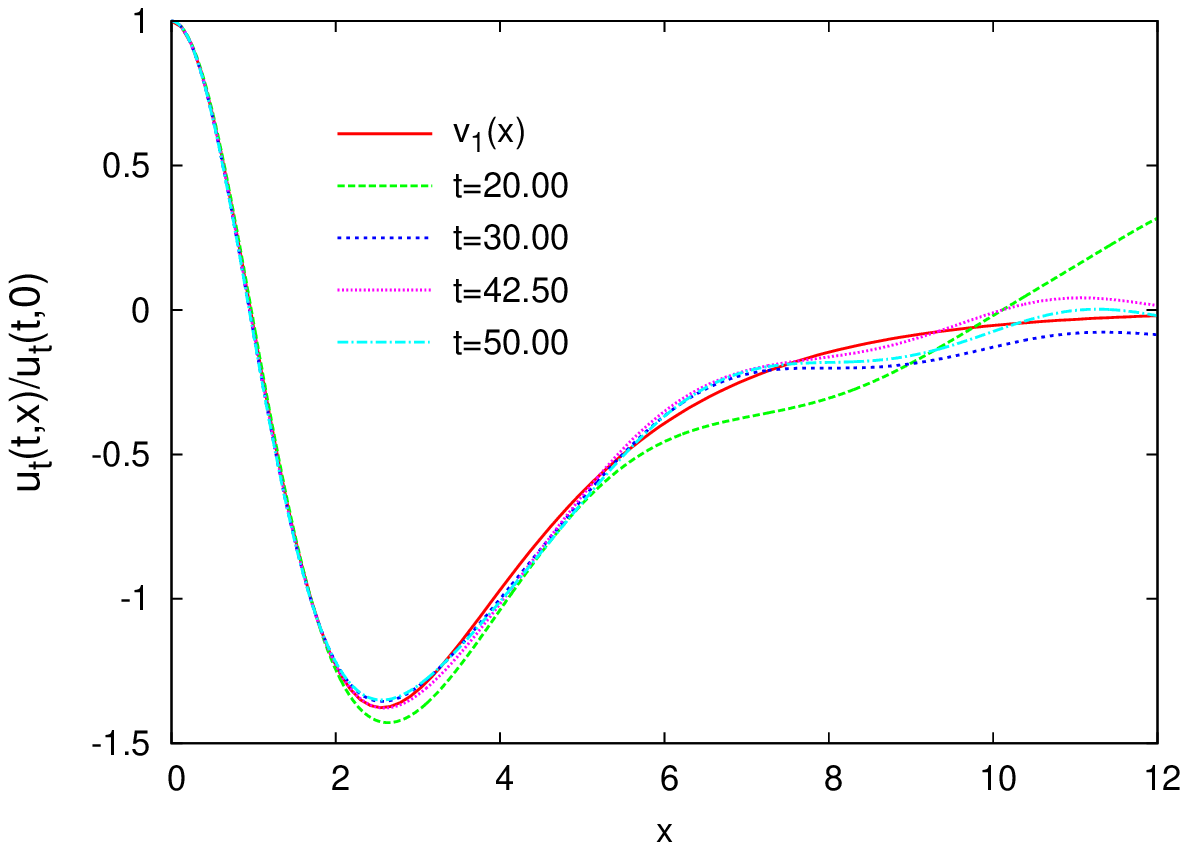}
\caption{$\alpha=1/2$. (a) Long-lived oscillation around $S$, (b) Convergence of the spatial profile to the oscillatory eigenmode (45).
\label{fig:p4}}
\end{figure}
\end{itemize}

The above results confirm that the near-critical dynamics is well described by the linearized approximation \eqref{lin_expansion}.
If there are no oscillatory modes (which happens for $\alpha\geq 1$), then we observe asymptotic convergence to $S$  in the form of an algebraic oscillatory tail with frequency $1$ and the decay rate
$\gamma=3/2$ for $\alpha>1$  and  $\gamma=1/2$ for $\alpha=1$. The spatial profile of the tail is shown to converge to the zero energy eigenfunction.
We emphasize that the oscillatory tail should not be confused with the so called quasinormal ringing which is governed by a different mechanism \cite{brz}.

 For $\alpha<1$ the late-time dynamics is dominated by the oscillatory modes. At the linear level the critical solution is asymptotically periodic (if there is only one oscillatory mode, which happens for $1/3\leq \alpha<1$) or quasiperiodic (if there are two or more oscillatory modes, which happens for $0<\alpha<1/3$). We expect that the nonlinearity  will induce a slow decay of the oscillatory modes via the resonant transfer of energy to radiative modes, however the calculation of this nonlinear decay is not an easy task, both numerically and analytically. Numerically, because
 even for very accurate bisection with $\delta\sigma\approx 10^{-30}$, the "lifetime" of near-critical dynamics is
  much too short to perform a reliable measurement of decay of amplitudes of oscillatory modes. Analytically, because the linear dispersive decay seems to be too weak to allow the calculation of radiation damping along the lines of \cite{sw}. Note that for $\alpha=1/2$ we have $2\omega_1>1$, hence the frequency $2\omega_1$  lies in the continuous spectrum, which indicates that  the nonlinear decay should occur already at the second order perturbation level. In Fig.~5 we verify that the deviation from the linearized approximation does in fact oscillate with frequency $2\omega_1$.

\begin{figure}[ht]
\center
\includegraphics[width=0.69\textwidth]{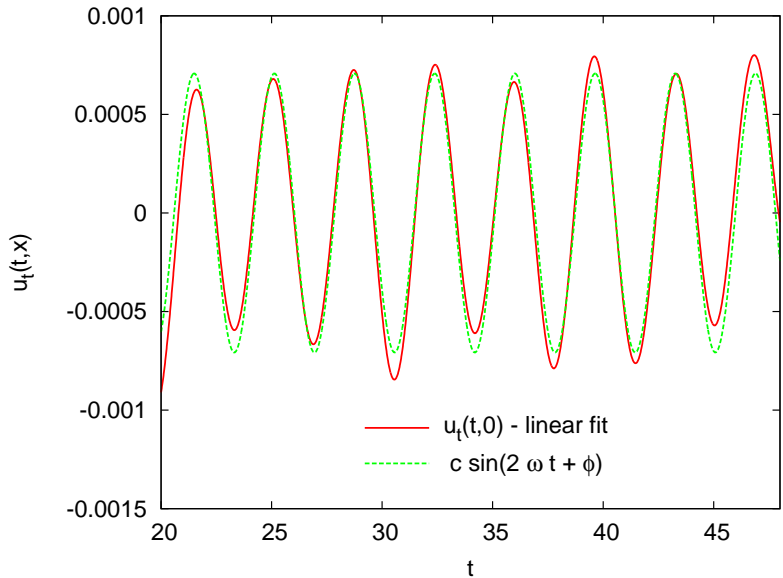}
\caption{$\alpha=1/2$. The deviation from the linearized approximation (obtained by subtracting the linear fit from the numerical solution) is shown to oscillate with the frequency $2\omega_1$.
\label{fig:p5}}
\end{figure}

\section{Final remarks}

 We hope that the explorations of the asymptotic behaviour of critical solutions described in this paper will inspire other researchers to carry on the investigation and prove our conjectures. We conclude with several remarks.
  \begin{enumerate}
 \item  In this paper we considered only even initial data. This restriction considerably facilitates the analysis since it eliminates  translation and boost symmetries and prevents the solution $S$ to move. For general data these symmetries must be accounted for by the modulation analysis. We believe that this is only a technical difficulty and
    our results for even data hold true in the general case provided that the assertions about the conditional asymptotic stability of $S$ are replaced by the conditional orbital asymptotic stability (with suitably red-shifted frequencies of oscillatory tails and modes) (cf. \cite{ns1,ns2} for the related problem in three space dimenions).
   \item We take this opportunity to point out that the asymptotic behaviour of solutions of Eq.\eqref{main} which decay to zero is not quite well understood.
         For $\alpha>1$ the globally regular small data solutions are asymptotically free for $t\ra\infty$, however
 for $\alpha\leq 1$ the asymptotic effect of the nonlinearity cannot be neglected and the solutions do not scatter to  free solutions \cite{g}.
  In the latter case one has to analyze the long range modulation of solutions, which (to our knowledge) has been accomplished only for $\alpha=1$ \cite{d}.
  Without the smallness assumption, as far as we know, the dynamics of globally regular solutions  in one space dimension is an uncharted territory (in higher dimensions the dispersion is stronger and the problem becomes easier, cf. \cite{imn,ksv}). For instance, it would be interesting to explore the post-ejection dynamics of marginally subcritical solutions and see how they enter into the small data  regime.

  \item
 As we saw above, perturbing the NLW equation by the mass term significantly affects the dynamics of globally regular solutions. In contrast, the asymptotics of blowup is structurally stable under this perturbation. Both for the NLW and the NLKG equations the blow-up is governed by the ordinary differential equation $u_{tt}-|u|^{2\alpha} u =0$, hence $u\sim (T-t)^{-1/\alpha}$ as $t\nearrow T$. This was first proved for the NLW equation by Merle and Zaag \cite{mz} and recently generalized to a class of perturbed NLW equations (including NLKG) by Hamza and Zaag \cite{hz} (in one space dimension these results hold for all $\alpha>0$).
\end{enumerate}
\vskip 0.2cm \noindent \textbf{Acknowledgments:}
The authors are grateful to the Erwin Schr\"odinger Institute  in Vienna, where part of this
work was done in February 2010 during the workshop "Quantitative Studies
of Nonlinear Wave Phenomena". PB acknowledges an inspiring discussion with Wilhelm Schlag and thanks Frank Merle for pointing out reference \cite{hz}.


\begin{thebibliography}{9}

\bibitem{c} T. Cazenave, \emph{Uniform Estimates for Solutions of
Nonlinear Klein-Gordon Equations}, J. Funct. Analysis 60, 36-55 (1985)

\bibitem{l} H. Levine, \emph{Instability and nonexistence of global solutions to nonlinear wave equations of the form $Pu_{tt} = - Au + \mathcal{F}(u)$}, Trans. Amer. Math. Soc. 192, 1–21 (1974)

\bibitem{kns} J. Krieger, K. Nakanishi, W. Schlag,
    \emph{Global dynamics above from the ground state energy for the one-dimensional NLKG equation}, preprint arXiv:1011.1776 [math.AP]

\bibitem{nist} F.W. Olver et al., NIST Handbook of Mathematical Functions, Cambridge University Press, 2010.

\bibitem{ps} L.E. Payne, D.H. Sattinger, \emph{Saddle points and instability of nonlinear hyperbolic equations}, Israel J. Math. 22, 273-303 (1975)

\bibitem{ds} R. Donninger, W. Schlag, \emph{Numerical investigation of the finite time blowup/global existence dichotomy for the cubic Klein-Gordon equation in $R^3$}, arXiv:1011.2015 [math.AP]

\bibitem{brz} P. Bizo\'n, A. Rostworowski, A. Zenginoglu, \emph{Saddle-point dynamics of a Yang-Mills field on the exterior Schwarzschild spacetime},  Class. Quantum Grav. 27, 175003 (2010)

\bibitem{sw} A. Soffer, M.I. Weinstein, \emph{Resonances and radiation damping in Hamiltonian partial differential equations},
Invent. Math. 136, 9–74 (1999)

\bibitem{ns1}  K. Nakanishi, W. Schlag, \emph{Global dynamics above the ground state energy for the focusing nonlinear Klein-Gordon equation}, arXiv:1005.4894 [math.AP]

\bibitem{ns2} K. Nakanishi, W. Schlag, \emph{Global dynamics above the ground state energy for the focusing nonlinear Klein-Gordon equation without the radial assumption}, arXiv:1011.0132 [math.AP]

\bibitem{g} R.T. Glassey, \emph{On the asymptotic behaviour of non-linear wave equations}, Trans. Amer. Math. Soc. 183, 187-200 (1973)

\bibitem{d} J-M. Delort, \emph{Existence globale et comportement asymptotique pour l'\'equation de Klein-Gordon quasi lin\'eaire \'a donn\'ees petites en dimension 1}, Ann. Sci. \'Ecole Norm. Sup. 34, 1-61 (2001)

\bibitem{imn} S. Ibrahim, N. Masmoudi, K. Nakanishi,
\emph{Scattering threshold for the focusing nonlinear Klein-Gordon equation},
arXiv:1001.1474 [math.AP]

\bibitem{ksv} R. Killip, B. Stovall, M. Visan,
\emph{Scattering for the cubic Klein--Gordon equation in two space dimensions},
arXiv:1008.2712 [math.AP]

\bibitem{mz} F. Merle, H. Zaag, \emph{Existence and universality of the blow-up profile for the semilinear wave equation in one space dimension}, J. Funct. Anal. 253, 43–121 (2007)
    
\bibitem{hz} M. Hamza, H. Zaag, \emph{A Lyapunov functional and blow-up results for a
class of perturbed semilinear wave equations}, arXiv:1002.2328v1 [math.AP]    
    



\end{thebibliography}
\end{document}